\documentclass[aps,pra,amsfonts,longbibliography,superscriptaddress]{revtex4-2}
\usepackage{bm,amsmath,amssymb,graphicx,tikz,epstopdf}

\usepackage{newtxtext}
\usepackage{newtxmath}
\usepackage{natbib}
\usepackage{hyperref}
\usepackage{xcolor,subeqnarray}
\hypersetup{
    colorlinks = true,
    urlcolor   = magenta,
    citecolor  = magenta,
}
\newcommand{\Sp}{\text{Sp}}

\begin{document}
	\title{Eukaryotic swimming cells are shaped by hydrodynamic constraints}

\author{Maciej Lisicki}

\email[]{mklis@fuw.edu.pl}

\affiliation{Faculty of Physics, University of Warsaw, Warsaw, Pasteura 5, 02-093 Warsaw, Poland}

\author{Marcos F. Velho Rodrigues}

\affiliation{Department of Applied Mathematics and Theoretical Physics, Centre for Mathematical Sciences, University of Cambridge, Wilberforce Road, Cambridge CB3 0WA, United Kingdom}

\author{Eric Lauga}

\email[]{e.lauga@damtp.cam.ac.uk}

\affiliation{Department of Applied Mathematics and Theoretical Physics, Centre for Mathematical Sciences, University of Cambridge, Wilberforce Road, Cambridge CB3 0WA, United Kingdom}

\begin{abstract}
	
	Eukaryotic swimming cells such as spermatozoa, algae or protozoa use flagella or cilia to move in viscous fluids. The motion of their flexible appendages in the surrounding fluid induces propulsive forces that balance with the viscous drag on the cells and lead to a directed swimming motion. Here, we use our recently built database of cell motility (BOSO-Micro) to investigate the extent to which the shapes of eukaryotic swimming cells may be optimal from a hydrodynamic standpoint. We first examine the morphology of flexible flagella undergoing waving deformation and show that their amplitude-to-wavelength ratio is near the one predicted theoretically to optimise the propulsive efficiency of active filaments. Next, we consider ciliates, for which locomotion is induced by the collective beating of short cilia covering their surface. We show that the aspect ratio of ciliates are close to the one predicted to minimise the viscous drag of the cell body. Both results strongly suggest a key role played by hydrodynamic constraints, in particular viscous drag, in shaping eukaryotic swimming cells.

\end{abstract}

\date{\today}

\maketitle

\section{Introduction}\label{sec:intro}

The cellular scale is home to a wide variety of biological swimmers, both in the prokaryote and eukaryote domains~\cite{yates86}. Examples that have long been studied comprise the spermatozoa of animals (including humans)~\cite{gaffney11}, algae and planktonic aquatic organisms~\cite{pedley92,stocker12,goldstein:15}, protozoa~\cite{jahn72} and bacteria~\cite{lauga16}. Prokaryotic swimmers exploit the rotation of passive helical filaments to self-propel~\cite{BergAnderson1973}.  In contrast, swimming eukaryotes employ a more complex propulsion machinery in which  flexible flagella actuated internally by molecular motors deform periodically in a wave-like motion~\cite{braybook}.

During recent decades, the biophysical and fluid mechanics communities have devoted considerable effort to quantify the interactions between cells and surrounding fluids~\cite{lighthill75,lighthill76,lauga_book}. Most studies use mathematical, or computational, modelling to rationalise the dynamics of cells, from individual motion in bulk fluids to collective interactions in complex environments~\cite{lp09,koch11,elgeti2015physics}. These theoretical tools have in turn allowed to   interrogate   the impact of fluid forces on shaping the geometry and kinematics of swimming cells, allowing  in particular to  pose cellular swimming as an optimisation problem, with notable successes. Recent work considered the impact of other biological constraints on bacterial shape~\cite{schuech2019motile}.

In the case of bacteria, in search of an optimal prokayotic flagellum, \citet{purcell97}  determined using simple arguments the shape of the most efficient helical propeller. A later theoretical study of the polymorphic forms of the flagellum commonly found in most bacteria by \citet{spagnolie_helix} showed that the so-called `normal' helical form is  the most efficient by some significant margin, suggesting that fluid mechanical forces may have played a  role in the evolution of the flagellum. In addition to the flagellum, overall swimming efficiency is also impact{ed} by the shape of the cell body,  as first shown and optimised by \citet{fujita01} for a monotrichous bacterium.

In eukaryotic cells, optimal propulsion strategies result from the interplay between fluid dynamics, elasticity and active driving mechanisms. In a study of over 400 mammalian species, \citet{tam2011optimal_2} discovered that spermatozoa have an optimal tail-to-head length ratio and that their most efficient swimming mode involves symmetrical, non-sinusoidal travelling waves in the tail. Such travelling waves have been shown in many contexts to lead to optimal motion in an active flagellum~\cite{pironneau74,pironneau_inbook,lauga2020traveling}. \citet{lighthill75} found the two-dimensional hydrodynamically optimal travelling wave profile to be sawtooth-like, but the singular kinks are regularised by elastic stresses~\cite{Spagnolie2010} or by properly modelling the irreversible work of molecular motors within the active flagellum~\cite{laugaeloy13}.
In ciliated organisms, the beating pattern of individual cilia has also been shown to be optimal for propulsion~\cite{osterman2011finding,eloylauga12}.  However, flagellar gaits may have local optima, suited e.g.~to distinct feeding and swimming modes, as investigated by \citet{tam2011optimal} for biflagellated phytoplanktons.  In larger organisms,   multi-ciliated cells    can increase their efficiency by choosing an optimal number of cilia   (from hundreds to thousands) \cite{omori2020swimming} and  both optimal swimming and feeding have been shown to involve surface metachronal waves~\cite{michelin2010,michelin2011}.

Beyond biological systems, the question of optimality has received a lot of attention  in the realm of biomimetic and engineered artificial swimmers. For example, swimming gaits and shapes maximising swimming efficiency have been determined for simple swimmers with only a few degrees of freedom~\cite{tam07,alouges08,golestanian08,AlougesDeSimoneLefebvre2009,nasouri2019efficiency}, two-dimensional cell shapes~\cite{avron04:opt,montenegro2014optimal} and idealised kinematics of treadmilling~\cite{leshansky07} {or surface slip flow driven swimmers~\cite{vilfan2012optimal}.  Other studies have sought shapes that would minimise absolute dissipation~\cite{nasouri2021minimum}.}

In this paper, we use data analysis to discover  remarkable optimal properties of the geometry of  cell shapes. We rely on our recently-assembled BOSO-Micro database~\cite{BOSO}: a comprehensive collection of data from the experimental literature on cellular swimming that contains data on swimming speed and morphological characteristics of 382 unicellular organisms, including spermatozoa. Building on the data, we first demonstrate  that the amplitude-to-wavelength aspect ratios of flagellated eukaryotes, including spermatozoa, are very close to the optimum predicted by hydrodynamic theory.
Next, we demonstrate that the aspect ratios (length-to-width) of ciliate cell bodies are well predicted by the classical result for the shape of minimum drag force subject to a fixed-volume constraint.

\section{Results}

The BOSO-Micro database~\cite{BOSO} provides us with novel means to examine swimming and morphological relationships across species. The database contains open-source swimming and morphological data published in scientific literature to date; we refer to the original paper for all details on data-gathering. Apart from swimming speed, many past investigations also contain data on {the basic morphology of organisms}, such as the size and shape of the cell body, or the length and number of flagella or cilia. The measurements and characterisation of organisms have been performed under different physiological conditions and in various environments, and therefore they do not represent a uniform sample. In cases where multiple values were reported for the morphology of swimming cells, we used average values.

\subsection{Propulsion of Flagellated Eukaryotes}

\begin{figure}
\centerline{\includegraphics[width=\linewidth]{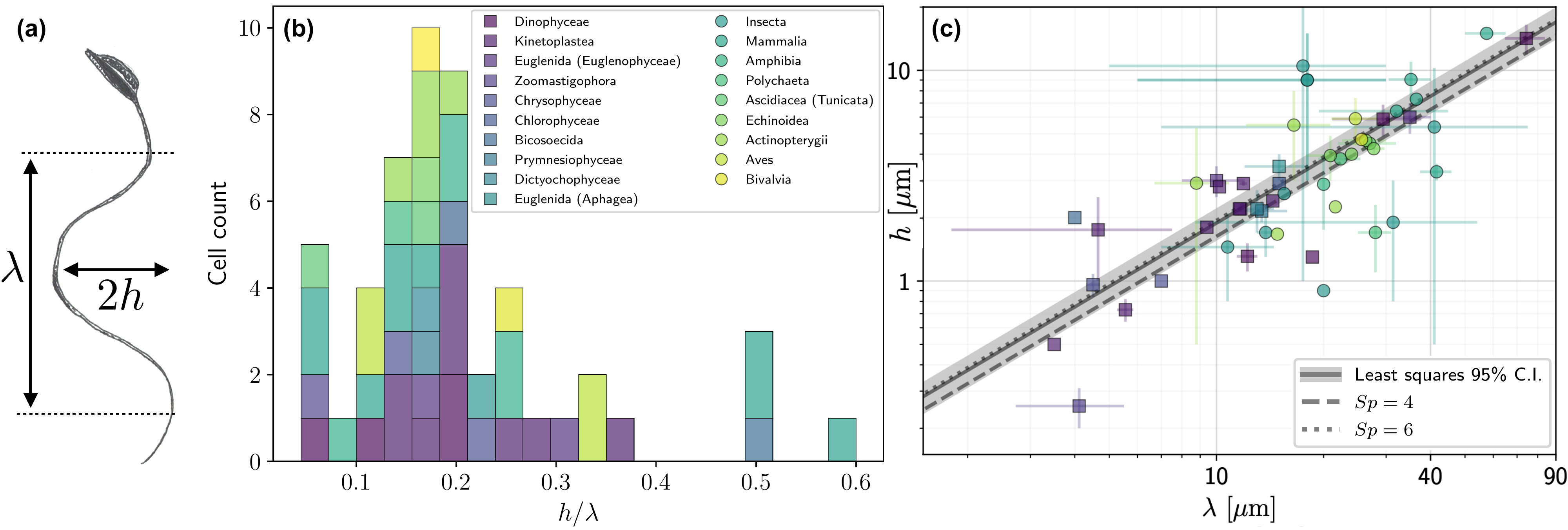} }
  \caption{{\bf Aspect ratio of eukaryotic flagellar waves}. (a): Sketch of flagellated eukaryotic cell with wave amplitude $h$ and wavelength $\lambda$. (b): Histogram of aspect ratios $h/\lambda$ from experiments gathered in  the BOSO-Micro database~\cite{BOSO}.
  (c): Mean  value of $h$ vs.~$\lambda$ for all  flagellated eukaryotes in our database. Spermatozoa are marked with circles, while the remaining flagellated eukaryotes are marked with squares. The solid  line  shows the least-square fit to the data and  the  shaded area is   the 95\% confidence interval.
  The dashed line shows the prediction from the optimal theory from~\citet{laugaeloy13} for $\Sp=4$ and the dotted one for $\Sp=6$. Cells from the same class in the tree of life are plotted using the same colour scheme, with all classes represented here listed in the inset of (b).}
\label{fig:fig1}
\end{figure}

 First, we focus on studies providing the details of actuation in flagellated eukaryotes and spermatozoa. We assume that the waveform of their flagellar oscillations can be captured by a characteristic amplitude $h$ and wavelength $\lambda$, as sketched in Fig.~\ref{fig:fig1}(a). The database contains 23 different species of flagellated eukaryotes and 28 different species of spermatozoa for which we found measurements of $h$ and $\lambda$; note that these values were either directly reported in each paper, or were estimated by us using experimental images, see details in~\cite{BOSO} .

For all our data, we plot in Fig.~\ref{fig:fig1}(b) the histogram of the amplitude-to-wavelength aspect ratio, $h/\lambda$. This ratio does not exceed 0.6 in the data set, with most organisms showing the value below 0.25; hence the amplitude of a flagellar beat rarely exceeds a quarter of the wavelength. 
 For further insight, we plot the amplitude $h$ against the wavelength $\lambda$ in Fig.~\ref{fig:fig1}(c). When multiple values are inferred for the same species, we plot in Fig.~\ref{fig:fig1}(c) the average value but include error bars to reflect the variability within the data {reported} for each species. Square markers represent flagellated eukaryotes, which are generally smaller in size as compared to spermatozoa,  marked with circles.

It is clear from the data in Fig.~\ref{fig:fig1}(c) that there is a  systematic increase of flagellar amplitude with wavelength. To rationalise and interpret these results, we performed a least-squares linear fit of the {$h/\lambda=\text{constant}$ relationship to the data.} The solid line shows the result of fitting, with the 95\% confidence interval shaded in grey. For comparison with theoretical predictions, we use the work of~\citet{laugaeloy13} who computed numerically the optimal shape of an active flagellum.  They considered an internally forced, elastic planar flagellum deforming periodically and moving at a constant speed through a viscous fluid and found the shape of the flagellum that maximises the swimming speed at a fixed energetic cost, representing irreversible work of molecular motors. They concluded that the optimal shape of the beating flagellum depends on a single dimensionless combination of parameters, termed the Sperm number $\Sp$~\cite{lauga_book}, and defined as the ratio of the wavelength $\lambda$ to the elasto-hydrodynamic penetration length $\ell$,
\begin{equation}
    \Sp = \frac{\lambda}{\ell}, \qquad\text{with}\quad \ell = \left(\frac{T B}{\zeta_\perp}\right)^{1/4},
\end{equation}
where $T$ is the period of oscillatory actuation, $B$ the bending rigidity of the flagellum, and $\zeta_\perp$ the perpendicular resistance coefficient of the flagellum (i.e.~the hydrodynamic force exerted on the flagellum per unit length for motion in the fluid perpendicular to its centreline). Thus the flagellum shape results from an interplay between the stiffness of the filament, viscous drag of the medium and internal forcing. { Accordingly, decreasing the value $\Sp$ results in flatter, smoother kinks in the flagellar beating \cite{laugaeloy13}.}

The typical values of the dimensionless $\Sp$ number for eukaryotes and spermatozoa have been report{ed} to be within the range of 1 to 10~\cite{Kumar2019}. For sea-urchin \textit{Arbacia punctulata} spermatozoa, \citet{Pelle2009} reported a bending rigidity of $B\approx 0.4 - 0.9\times10^{-21}$~Nm$^2$, while \cite{howard01} assume it to be the upper limit of that range. In direct measurements, \cite{okuno79} found the stiffness of echinoderm sperm flagella to be in the range of $B\approx 0.3 - 1.5\times10^{-21}$~Nm$^2$, while for \textit{Strongylocentrotus purpuratus}, \cite{Okuno1981} estimated $B\approx 0.8\times10^{-21}$~Nm$^2$. Mechanical measurements for the flagella of wild-type {\it Chlamydomonas} yield $B = (0.84 \pm 0.28)\times10^{-21}$~Nm$^2$ \cite{xu2016flexural}, with similar values for mutant strains. In water, we have $\zeta_\perp\approx 2\times10^{-3}\ {\rm Pa\,s}$, and the typical periods $T$ vary across species from 0.02 to 0.05~s \cite{brennen77,Gray1955b,BOSO}. From these numbers, and typical lengths of dozens of $\mu$m \cite{BOSO}, we therefore estimate  the values of $\Sp$ to be in the range of $\Sp \approx 2 - 7$.

To compare the experimental measurements to predictions from theory, we use the work of \citet{eloylauga12} where, for  a given value of $\Sp$,   the shape of the optimal flagellum was computed;  for each optimal shape, we can then  extract the corresponding optimal aspect ratio, $h/\lambda$. In Fig.~\ref{fig:fig1}(c) we plot the aspect ratio as a linear curve $h/\lambda=constant$ for $\Sp=4$ ($h/\lambda=0.163$, dashed line) and $\Sp=6$ ($h/\lambda=0.194$, dotted line). Both lines lie within the 95\% confidence interval of our dataset: the best fit for the data yields an aspect ratio $h/\lambda=0.188$, with the lower and upper bound of the 95\% confidence interval at $0.163$ and $0.213$, respectively ($R^2=0.562$).

\subsection{Hydrodynamic drag force on ciliates}

\begin{figure}

\centerline{   \includegraphics[width=\linewidth]{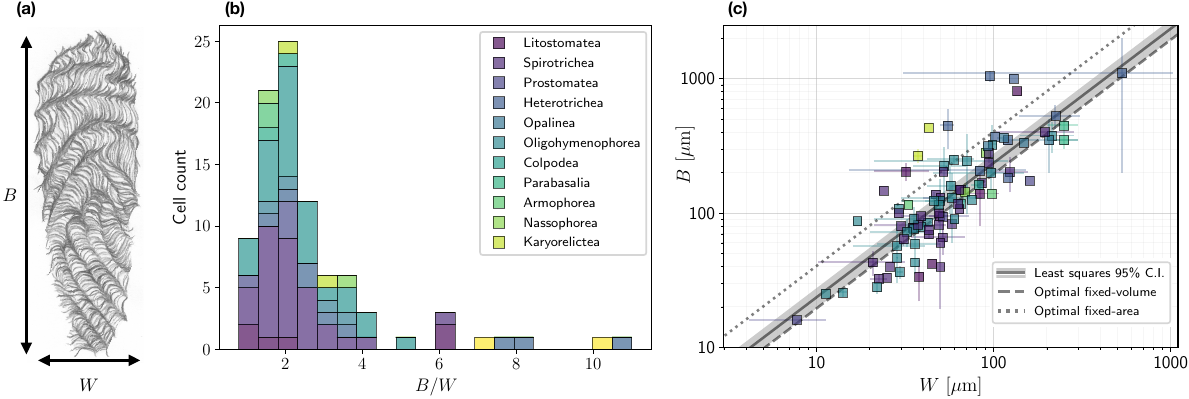}}
\caption{{\bf Aspect ratio of ciliates cell body}. (a): Sketch of ciliate body showing the cell width $W$ and length $B$.
  (b): Histograms  of cell aspect ratios $W/B$ from  experimental data gathered in the BOSO-Micro database~\cite{BOSO}.
  (c): Mean value of body length $B$ (in $\mu$m)   plotted as a function of  width $W$ (in $\mu$m) for all species in  our database. The  solid line  shows the least-square fit to the data  while  the   shaded  area is the 95\% confidence  interval. The  dashed line shows the prediction from  the  theory of fixed-volume drag  minimisation ($B/W=1.952$)~\cite{pironneau1973}
  while the dotted line the drag minimum for  a given surface area  ($B/W=4.037$)~\cite{montenegro2015other}. Cells from the same class in the tree of life are plotted using the same colour scheme, with all classes represented here listed in inset of (b).
  }
\label{fig:fig2}
\end{figure}

Ciliates constitute a large group of unicellular eukaryotic swimmers, characterised by the presence of hair-like cilia which cover their bodies in dense active carpets; a model organism, {\it Paramecium}, is sketched in Fig.~\ref{fig:fig2}(a). The internal structure of cilia is identical to eukaryotic flagella but they are typically shorter and used to propel cells with relatively much larger body sizes. This separation of scales has motivated   mathematical models dating back to  \cite{Blake1971a} to represent the action of  cilia layers covering the cell bodies as a continuum boundary with prescribed actuation of the surrounding fluid. This surface forcing sets the fluid in motion, and leads to swimming. The balance of propulsion force and viscous drag force on the cell body determines the swimming speed $U$~\cite{lauga_book}. Because the latter results from the overall shape of the body, one possibility for cells to increase their swimming speed is to reduce the viscous drag opposing their propulsion by choosing the appropriate drag-minimising shape, which have been characterised theoretically~\cite{pironneau1973,pironneau1974optimum,montenegro2014optimal}.

To investigate if cells aligned with expectations from drag-minimisation, we use our eukaryotic microswimmers database~\cite{BOSO} to quantify the extent to which the cell bodies of ciliates minimise drag forces.  In Fig.~\ref{fig:fig2}(b), we plot the histogram for the aspect ratio of the cell bodies, $B/W$, of the 91 ciliates in our database. When multiple values were available for the same species, we use their average. Colours are used to code different classes of organisms. Clearly, the reported cells are typically prolate, and most of them have a length of about twice their width, although more slender cell bodies are also observed.

To further explore the correlation between the two length scales characterising the  cell bodies of ciliates,  we present in Fig.~\ref{fig:fig2}(c) a scatter plot of the data, which suggests a linear correlation between cell body and width.  We plot the linear least-square fit to the data with a solid line, corresponding to the slope   $B/W=2.360$ ($R^2=0.503$), while grey shading marks the 95\% confidence interval, for aspect ratios in the range $(2.057,\ 2.664)$. Remarkably, these results  are very close to the classical prediction of a minimal-drag shape in Stokes flow by \citet{pironneau1973}, later computed by \citet{Bourot1974}, who demonstrated numerically that the aspect ratio  {$B/W=$}~1.952 is hydrodynamically optimal for bodies with prescribed volumes (the corresponding optimal shape resembles a prolate spheroid with pointy ends). We mark this theoretical prediction with a dashed line in Fig.~\ref{fig:fig2}(c), and note that it lies just outside the confidence interval of our fit to data.     Note that, in contrast, if instead one considers the   shape that has minimum drag for a  prescribed surface area (instead of a fixed volume), \citet{montenegro2014optimal} found the optimal aspect ratio to be 4.037, marked by in Fig.~\ref{fig:fig2}(c) with the dotted line; this lies well outside the data.

It is worth noting that the Stokes drag acting on an optimal fixed-volume prolate shape is only 4.5\% less than that experienced by a sphere of the same volume, suggesting that the energetic cost of diverting from the optimum is actually relatively low. It is therefore perhaps all the more remarkable that such a small energetic gain seems to be reflected in the empirical data. The small magnitude of the energetic improvement might in turn allow a larger cell shape variability and explain the diversity of observed micron-scale shape features.

\section{Discussion}

In this paper, we used morphological data gathered for swimming eukaryotic microorganisms in the Micro-BOSO database  which assembled published data from the literature \cite{BOSO} to test two outstanding questions regarding hydrodynamic optimality of the geometric characteristics and shapes of the cells.  First, we have shown that for 51  different species of flagellated unicellular eukaryotes and spermatozoa, which had the characteristics of their flagellar waves measured and reported in the literature, the ratio of the wave amplitude to the wavelength was close to the optimal hydrodynamic predictions  for the shape of energetically-optimal active filaments  for Sperm numbers comparable to estimates for natural swimmers.  Second, we have analysed a dataset of body widths and lengths of 91 species of ciliates and   demonstrated that their cells have an aspect ratio close to the theoretical optimum for the minimal drag on a body of a fixed volume.

Our findings  shed new light on the question of energetics of microscale swimming. There  is a long-standing argument in biophysics that the energetic cost of locomotion is negligible when compared with metabolic constraints~\cite{purcell77}. Recent work focusing on   169 species of eukaryotes  has estimated that the operating costs of flagella  remain small, at a few percent of the total energy budget, but flagellar construction costs can be quite significant~\cite{schavemaker2022flagellar}. The results in the current work strongly indicate that, in multiple species of  swimming cells that span  almost three orders of magnitude in length, cellular shapes are consistent with energetic optimisation. This remarkable agreement suggests that hydrodynamic forces  might have played a key role in shaping biological active swimmers.

Clearly, not all swimmers in Figs.~\ref{fig:fig1} and \ref{fig:fig2} follow exactly the theoretical curves. Outliers in Figs.~\ref{fig:fig1} that are the furthest from the best-fit
line are  \textit{Giardia lamblia}, Cricket, Worm, Midge and Rabbit spermatozoa (below the trend line), and \textit{Poteriodendron}, Stick insect and Fly spermatozoa (above the line). In Fig.~\ref{fig:fig2} the outliers are \textit{Halteria grandinella}, \textit{Mesodinium rubrum} (below the line) and \textit{Spirostomum} spp.~and \textit{Trachelocerca} spp.~(above the line). Upon examining  the shape of these specific organisms we could not identify any systematic distinguishing features. Other factors are therefore at play in determining the morphology of  some species. In the case of bacteria, it is   known that cell shapes are the result of a delicate trade-off between efficient swimming, the chemotaxis budget, and  the costs related to cell construction~\cite{schuech2019motile}, which points to the complexity of the overall energetic landscape.

 Furthermore, in the case of ciliated organisms, minimising the drag force might not be sufficient for maximising propulsion, since swimming is a balance between drag and thrust. And yet, the available experimental data strikingly show  that drag minimisation appears to be  important for ciliates, independently of the details of their propulsion efficiency. We conjecture that  this might stem from the necessity of these cells to not only self-propel but  also to respond  to the external environmental flows to which they are invariably exposed.

\section*{Acknowledgements}

This project has received funding from the European Research Council (ERC) under the European Union's Horizon 2020 research and innovation programme  (grant agreement 682754 to EL) and from the National Science Centre of Poland (grant Sonata no. 2018/31/D/ST3/02408 to ML) and from Campus France (Eiffel Scholarship no. 812884G to MFVR). 

\section*{Data availability} 

The data that support the findings of this study are openly available in the OSF repository \url{https://osf.io/4tyx6}. It is also available and editable on GitHub: \url{https://github.com/marcos-fvr/BOSO-micro}.

\end{document}